\documentclass[twoside]{article}
\usepackage[utf8]{inputenc}
\usepackage[square,sort,comma,numbers]{natbib}
\usepackage{graphicx}
\usepackage{ragged2e}
\usepackage{changepage}
\usepackage{csquotes}
\usepackage{amsmath}
\usepackage{hyperref}
\usepackage{xcolor}
\usepackage{fancyhdr}
\usepackage{fourier-orns}
\usepackage{amssymb}
\usepackage{dsfont}
\usepackage[top=4cm,bottom=4.5cm,left=4.7cm,right=4.7cm,asymmetric]{geometry}
\definecolor{colour}{HTML}{0645AD}
\hypersetup{
    colorlinks=true,
    linkcolor=black,
    citecolor=black,
    urlcolor=colour,
}
\emergencystretch=\maxdimen
\begin{document}\thispagestyle{empty}

\begin{center}
    \Large\textbf{Reversible Colour Density Compression of Images using cGANs}
\end{center}
\begin{center}
    \vspace{4px}
    \large\textbf{Arun Jose} and \textbf{Abraham Francis}
\end{center}
\begin{center}
    \vspace{-10px}
    jozdien@ieee.org, abraham@cet.ac.in
\end{center}

\begin{center}
    \vspace{35px}
    \large\textbf{Abstract}
\end{center}

\begin{adjustwidth}{1cm}{1cm}
\justify{
    Image compression using colour densities is historically impractical to decompress losslessly.  We examine the use of conditional generative adversarial networks in making this transformation more feasible, through learning a mapping between the images and a loss function to train on.  We show that this method is effective at producing visually lossless generations, indicating that efficient colour compression is viable.
    }
\end{adjustwidth}

\tableofcontents

\section{Introduction}

Much of modern research into image compression involves making the compressed image visually lossless.  This led to the development of algorithms based on methodologies such as transform coding - examples include the Discrete Cosine Transform\cite{raid2014jpeg}, the wavelet transform\cite{jpeg-2000} and fractal compression.  With the advent of modern deep learning models however, there exists a broader scope for compression degrees and methods due to their ability to process decompression.

Another area of research in compression methods is based on the density and focal points of colour present in an image.  Examples include the study of chroma subsampling and the use of DCT algorithms\cite{abdelhafiez2012color}\cite{color-dct-table}.  These approaches differentially compress the vital parts of an image (high-quality) and the remainder (lower quality), to maximize compression for the least visual loss.  This is especially useful when de-compression is an impractical option for such transformations.

Modern deep learning networks, however, do provide such capacity for de-compression.  Convolutional neural nets (CNNs) have already seen usage in de-compression\cite{zhang2020nearlossless}.  CNNs are trained to minimize a specified loss function that provides a metric for the quality of results.  Though the training is automatic, there is a lot of work that goes into specifying what the loss is, which is where Generative Adversarial Networks (GANs)\cite{goodfellow2014generative}\cite{agustsson2019generative} come into play.  GANs have the quality of simultaneously learning the loss function and trying to minimize it.  This allows it applicability in a broad scope of functionalities.  Conditional GANs (cGANs) use labels as an extension to GANs to generate and discriminate images better\cite{mirza2014conditional}.  When conditioned on an input image and generating a corresponding output image, cGANs are suited to image-to-image translation\cite{pix2pix2017}.

In this paper, we examine image compression through reducing the colour density of an image (the number of bits required to encode one colour pixel), and reverting to a degree the de-compressed image using conditional GANs trained for translation between two kinds of images.  The cGAN architecture comprises a U-Net based architecture\cite{ronneberger2015unet}, and a convolutional PatchGAN classifier for the discriminator \cite{pix2pix2017}.  Code can be found \href{https://colab.research.google.com/drive/1eIQfcljLeq5gWOipm9fyuVoVvX0zenLZ}{here}.

\section{Method of Compression}

Images are compressed through a reduction in the colour density of each pixel.  In the images sampled, pixels are stored as 3-byte RGB compositions - i.e, 3*8 bits, representing $2^{8^{3}}$ tonal levels.  Compression is performed by collapsing this composition by reducing the number of bits (and subsequently, the number of tonal levels) stored in each part of the RGB composition.  This is represented by the formula
\[
x = \frac{8}{8 - b},
\]
where $x$ is the compression factor - the ratio of the sizes of the raw image and the compressed image, and $b$ is the number of bits dropped in each part of the composition.  

In this paper, we take samples of $b = 5$, dropping 15 bits from each pixel, giving a compression factor $x$ of 2.667, and $b = 4$, dropping 12 bits from each pixel, giving us a compression factor $x$ of 2.

There exist more visually lossless methods of colour-based compression\cite{abdelhafiez2012color}\cite{color-dct-table}; however, for the purpose of reducing the complexity of the decompression neural net architecture, a uniform scaling mechanism was chosen.  This has the added advantage of precise degrees of compression for any input file, providing more reliable results.

\section{Method of Decompression}

Conditional GANs learn a mapping from an input image $x$ and a random noise vector $z$ to the output image $y$ as $G: \{x, z\} \rightarrow y$, where the generator $G$ is trained to produce images that are realistic enough to be indistinguishable by a discriminator $D$, which is conversely trained to identify the generated images.

The loss function, as expressed in Isola et al., 2017 \cite{pix2pix2017}:
\[
\mathcal{L}_{cGAN}(G,D) = \mathds{E}_{x,y}[log D(x,y)] + \mathds{E}_{x,y}[log(1 - D(x,G(x,z))],
\]
\[
\mathcal{L}_{L1}(G) = \mathds{E}{x,y,z}[\lVert y - G(x,z)\rVert_1],
\]
The final objective thus described is:
\[
G^{*} = \arg\min_{G}\max_{D}\mathcal{L}_{cGAN} (G,D) + \lambda\mathcal{L}_{L1}(G)
\]

Justifications for this formalization can be found in the referred paper; the focus of this paper is a specific use case of this network.

\subsection{Network Architecture}

The generator and discriminator architectures are adapted from \cite{pix2pix2017}, and use modules of the form convolution-BatchNorm-ReLu.  Further details are outlined in that paper and its supplemental materials, with only key points of difference being described below.

\subsubsection{Generator}

The generator architecture follows the general structure of a U-Net\cite{ronneberger2015unet}.  Here, we use a double-pass U-Net with skip connections between the layers $i$ and $n - i$ where n is the total number of layers, for every $i$.  

The high-resolution input grid is converted into low-resolution intermediates twice in this structure, along with one high-resolution intermediary.  The image data is filtered through a bottleneck twice.  Thus, there are two downsampling and two upsampling stacks.

\subsection{Training Schema}

The model was trained on a custom dataset of 411 images, that can be found \href{https://www.kaggle.com/jozdien/pictures}{here}.  The model was tested on frames extracted from a short clip that were compressed.  Results chosen at random are shown in Figure 1.  Fragments where the model performed poorly are highlighted in Figure 2.  

\begin{figure}
    \centering
    \includegraphics[keepaspectratio, width=12cm]{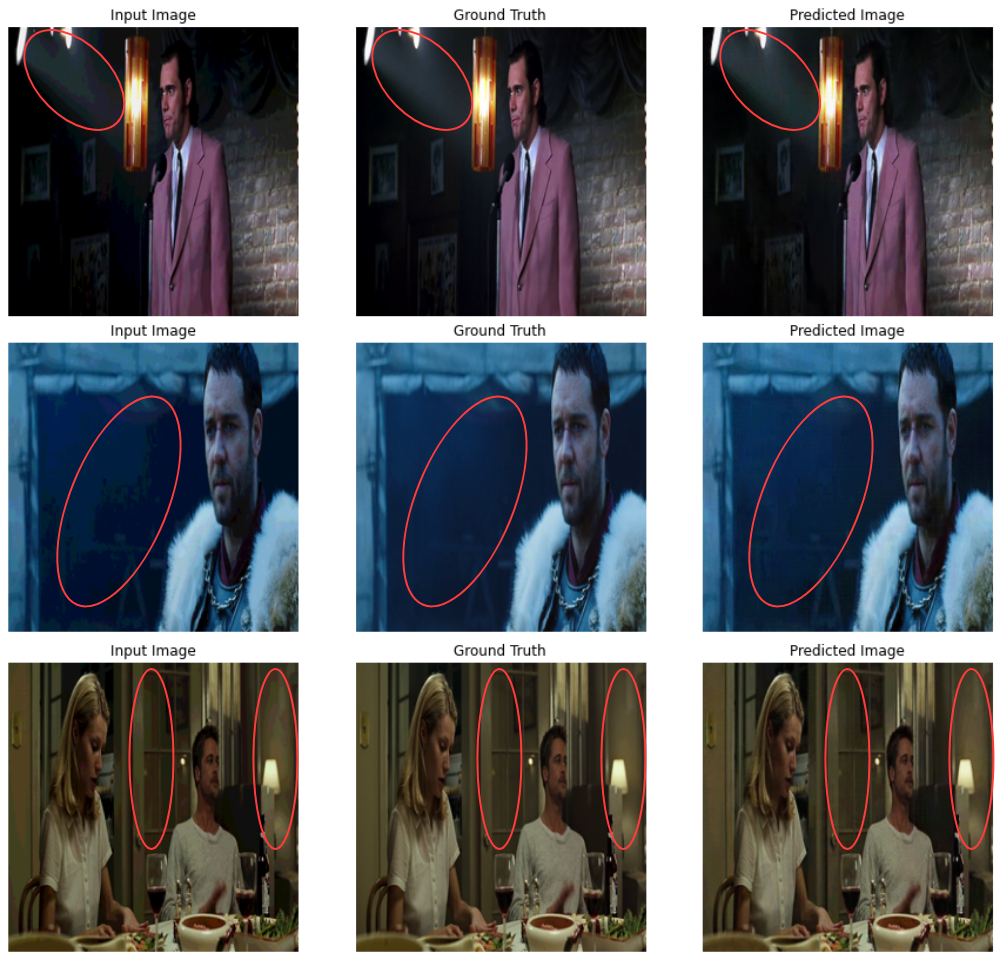}
    \caption{Sample results using movie stills.  Notice that gradient areas are the most affected by the compression, and are thus the focus of the comparison.}
    \label{fig:1}
\end{figure}

\begin{figure}
    \centering
    \includegraphics[keepaspectratio, width=12cm]{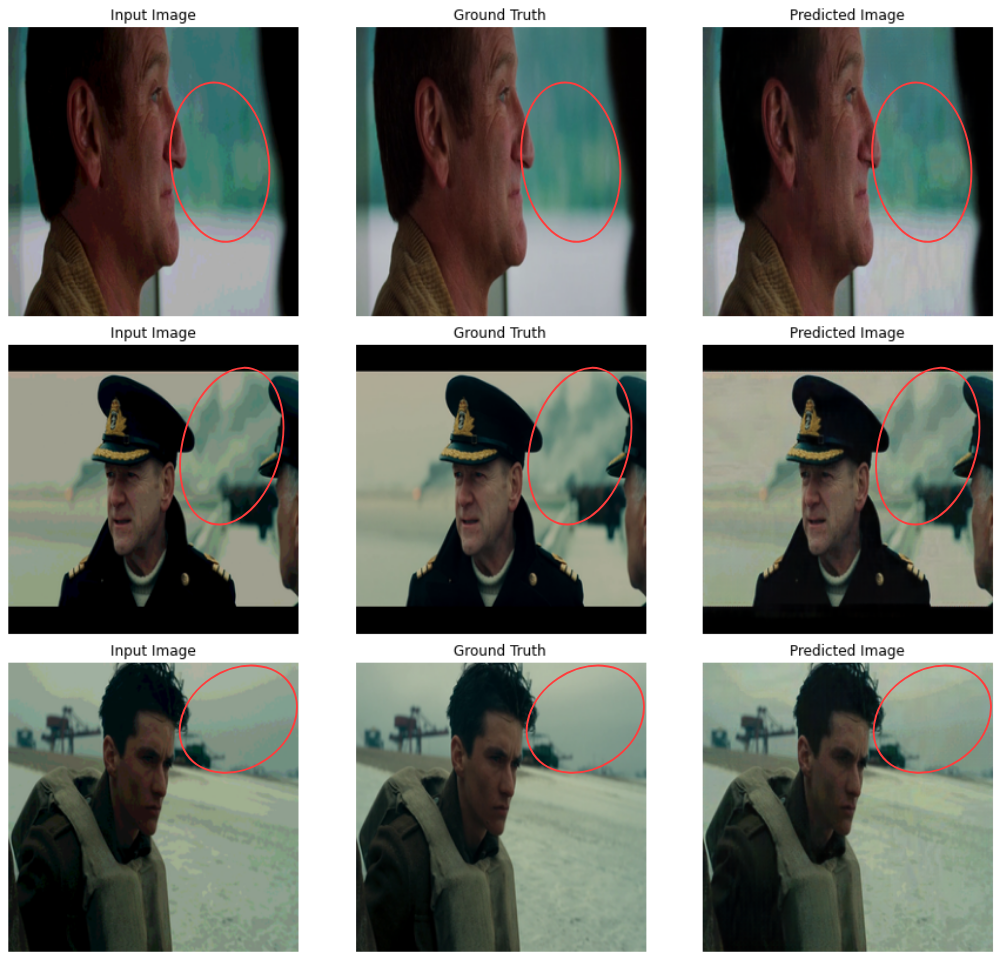}
    \caption{Sample failure cases.  Failures are generally concentrated in areas with high gradients - blurred regions, for example, which are fortunately areas of low user visual concentration.}
    \label{fig:2}
\end{figure}

Training was relatively fast for the size of the model.  For the results shown in the figures, training time was roughly two hours on an Nvidia Tesla T4 GPU.  During testing, processing takes less than a second per data item.

\newpage

\section{Testing}

Formally assessing the quality of generated images is a difficult problem\cite{salimans2016improved}.  We test whether the generated images can fool human judges, relative to the original.  For many uses such as in video streaming and graphics, this test of visual losslessness is the goal.

For the experiment, each judge was shown either an original video or the model-decompressed version at random.  They were allowed to view the video as many times as required, and then decide whether it was the original or the reconstructed.  The final metric would be how close the percentage of responses in either direction are for the original and generated videos.  

Of 55 people evaluated, the original video had evenly split responses, and the generated video had 48.8\% of responses for being the original, and 51.2\% for being generated.  Given the sample size, this similarity constitutes fair evidence for the visual losslessness of the generated video.

\section{Conclusion}

The results in this paper suggest that conditional GANs show promise as a relatively straightforward approach to a previously complex problem in colour-density reconstruction in decompression.  The mechanisms used in this work are prototypical, we hope that more theoretically-focused applications will achieve even better results.

\bibliographystyle{unsrtnat}
\bibliography{references}
\end{document}